\documentclass[runningheads]{llncs}
\usepackage[T1]{fontenc}
\usepackage{graphicx}
\usepackage{hyperref}
\usepackage{color}

\usepackage[frozencache]{minted}
\usepackage{amsmath}
\usepackage{paralist}
\usepackage{enumitem}
\usepackage{dirtree}
\begin{document}
\title{More Powerful Constant Value Checking with SMT Solving}
\author{Jonas Mittnacht\inst{1}\orcidID{0009-0005-4030-0125} \and
	Florian Lanzinger\inst{1}\orcidID{0000-0001-8560-6324}}
\authorrunning{J. Mittnacht et al.}
\institute{Institute of Information Security and Dependability (KASTEL), Karlsruhe Institute of Technology, Karlsruhe, Germany\\
\email{jonas.mittnacht@student.kit.edu}}
\maketitle              % typeset the header of the contribution
\begin{abstract}
	Pluggable type systems extend the basic type system of a programming language by introducing additional type hierarchies to specify more complex properties and dependencies between variables.
    We consider the Constant Value Checker, a pluggable type checker for Java implemented using the \emph{Checker Framework}, which provides various pluggable type systems and corresponding checkers.
	The Constant Value Checker lets users specify restrictions on the values a primitive integer or boolean variable can hold using a set or range of constant values.
	However, the current implementation of the Value Checker makes use of a limited set of syntactic typing rules that, in practice, often fail to verify well-typedness of more complex code.

	In this paper, we introduce an extension to the Checker Framework using \emph{Satisfiability Modulo Theories (SMT) solvers} to address these limitations by creating corresponding first-order logic formulas for expressions whose type checks fail with the current syntactic typing rules and determining their well-typedness based on the satisfiability of these formulas.
	Furthermore, we introduce new \emph{dependent types} for specifying allowed values with expressions that can depend on other variables in the program.

\end{abstract}
\section{Introduction}
\paragraph{Background}
Pluggable type systems are type systems defined on top of a language's base types and allow the specification of additional properties.
To verify and strengthen guarantees about a Java program ahead of execution, the Checker Framework~\cite{checkerframework} project provides a set of such type systems and uses so-called \emph{Checkers} to verify the correct usage of these types.
One such pluggable type system in the Checker Framework is the Constant Value Checker~\cite{constvalchecker}.
It provides a set of annotations for denoting that the value of a variable of a primitive or \texttt{String} type (or the length of an array) is restricted to a set of constant values specified in the annotation.
However, the type-checking capabilities of the underlying Checker implemented in the framework are rather limited in power, particularly because it does not track correlations (e.g.~equality) between variables.
In expressions with strongly correlated variables, the Checker therefore often evaluates the resulting value to be within a much broader range than required - as an example, for an expression of the form \texttt{(x + 1) * (x - 1)} in which \texttt{x} is known to hold a value within the range $[-2, 2]$, the Constant Value Checker is unable to verify that the resulting value is indeed in the range $[-1, 3]$.
\paragraph{Contribution}
To alleviate these limitations, we present a method for validating constant value types using \emph{Satisfiability Modulo Theories (SMT}) solving.
We implemented this approach for the Checker Framework's value types in a fork%
\footnote{The implementation of this extension alongside a guide on how to run the Value Checker with the SMT extension can be found as an artifact on Zenodo~\cite{artifact} or in the development repository at \url{https://gitlab.kit.edu/uioee/checker-framework}.}
of the framework and compared it to the existing Value Checker w.r.t.\ its performance and its expressiveness.
\paragraph{Related work}
Our work is related to more general \emph{refinement type systems}.
This term was introduced by Freeman and Pfenning~\cite{freeman-1991} for a type system that refines algebraic data types by restricting the choice of constructor based on an expression's type.
Nowadays, the term generally refers to type systems that combine a base type (e.g., \texttt{int}) with a refinement formula that must hold for all instances (e.g., $-1 \leq x \leq 3$).
Most of these kinds of refinement type systems decide the well-typedness of a program by translating type constraints into formulas to be checked by an SMT solver or some other theorem prover~\cite{xi-1998,rushby-1998,rondon-2008,vazou-2014,toman-2020,LanzingerWeiglUlbrich2021_1000139256,gamboa-2023}.

\newcommand{\fol}[1]{\mathit{fol}_{#1}}
\newcommand{\conds}[1]{\mathit{conds}_{#1}}
\newcommand{\constrain}[1]{\mathit{constrain}_{#1}}
\newcommand{\folcontext}[1]{\mathit{ctx}_{#1}}
\newcommand{\foltypecontext}[1]{\mathit{typectx}_{#1}}
\section{Overview and Concept}
\label{sec:overview}
The aim of our extension is to alleviate the aforementioned limitations specifically for type checks on expressions of the Java integer (\texttt{byte}, \texttt{short}, \texttt{int}, \texttt{long} and \texttt{char}) and \texttt{boolean} types using SMT solving.
The Constant Value Checker has three existing value types for integer and boolean base types that specify allowed values using annotations:
\begin{itemize}
	\item \mintinline{java}{@IntVal({x1,...,xN})} for integer values in the set \texttt{\{x1,\,\dots,\,xN\}},
	\item \mintinline{java}{@IntRange(from=xL, to=xR)} for integer values in the range \texttt{[xL,\,xR]},
	\item \mintinline{java}{@BoolVal({x1,...,xN})} for boolean values in the set \texttt{\{x1,\,\dots,\,xN\}}.
\end{itemize}
Our extension introduces additional \emph{dependent} value types, which make it possible to specify the allowed values in relation to other expressions in the program.
These expressions of a dependent type are not evaluated during type checking -- instead, their translations are incorporated into the type checking formulas as either bounds for or potential values of variables of that type.%
\footnote{The expressions of a dependent type are incorporated as if they were evaluated just before the variable of that type occurs in an expression.}
To avoid issues with mutating values, we require these expressions to be constant within the scope they are declared in, by requiring all referenced variables to be \texttt{final} and disallowing method calls.
There are three new dependent types:
\begin{itemize}
	\item \mintinline{java}{@IntValExpr({e1,...,eN})}, which specifies a set of expressions \texttt{\{e1, \dots, eN\}} in \mintinline{java}{String} form that evaluate to an integer type, of which at least one must evaluate to the typed element's value,
          %\todo[inline]{Well-definedness. I.e., what if the expressions e1,...,eN aren't of type int or don't terminate or have side effects?}
    %
	\item \mintinline{java}{@IntRangeExpr(from=eL, to=eR)}, which specifies a lower and an upper bound using expressions \texttt{eL} and \texttt{eR}, respectively,
	\item \mintinline{java}{@BoolValExpr({e1, ..., eN})}, which does the same as \mintinline{java}{@IntValExpr(...)} for the \mintinline{java}{boolean} type.
\end{itemize}

To increase the power of the Value Checker, we reduce the correctness of a given value type $\tau_C$ on an expression $e$ in the program to the (un)satisfiability of a corresponding first-order logic formula which we call its \emph{type check formula}.
\begin{figure}
    \centering
    \begin{minipage}{0.8\textwidth}
        \begin{minted}[autogobble, linenos]{java}
        @IntRange(from=1, to=11) int foo(int x) {
            if (x > 10 || x < -10) { return 1; }
            if (x < 0) { x = -x; }
            return x + 1; }
        \end{minted}
    \end{minipage}
    \caption{Sample Java program for demonstrating type check formula creation.}
    \label{fig:type-check-formula-sample-prog}
\end{figure}
Consider the method \texttt{foo} in Figure \ref{fig:type-check-formula-sample-prog}.
The type check formula for the statement in line 4 would be of the form
\begin{align*}
             & \mathrm{res} \doteq x_1 + 1 \\
    {} \land & (x_0 \geq 0 \doteq \mathtt{true} \rightarrow (x_1 \doteq x_{d_1} \land x_{d_1} \doteq x_0)) \\
    {} \land & (x_0 < 0 \doteq \mathtt{true} \rightarrow (x_1 \doteq x_{d_2} \land x_{d_2} \doteq -x_0)) \\
    {} \land & (x_0 \geq -10 \doteq \mathtt{true} \land x_0 \leq 10 \doteq \mathtt{true}) \\
    {} \land & (\mathrm{res} < 1 \doteq \mathtt{true} \lor \mathrm{res} > 11 \doteq \mathtt{true}).
\end{align*}
This formula incorporates
\begin{enumerate}
    \item the translation $\mathrm{res} \doteq x_1 + 1$ of the type-checked statement.
    \item the exact value $x_1$ at the point the statement is executed of the variable \texttt{x} depending on its initial value $x_0$ and the preceding control flow ($x_1 \doteq x_0$ if $x_0 >= 0$, $x_1 \doteq -x_0$ otherwise).
    Note that we use a static single-assignment transformation, introducing separate logical variables $x_i$ for each assignment to the program variable \texttt{x}.
    For variables assigned in a loop or for fields that may be assigned unknown values by method calls, we cannot give an exact value in the formula; in these cases, we only use the variable's declared type, which must hold in every state.
    \item the path condition $x_0 \geq -10 \doteq \mathtt{true} \land x_0 \leq 10 \doteq \mathtt{true}$ under which the expression is evaluated, i.e. the conditions of conditional branches and loops in which the type-checked statement resides.
    %\todo[inline]{Loops? Either mention something here or in Sec. 3.2}
    %\todo[inline]{JM: Not sure what to include here - the condition guarding a loop itself can be incorporated just fine, and if it contains a variable that changes within the loop, we proceed as mentioned in the previous point.}
    %
    \item the condition $\mathrm{res} < 0 \doteq \mathtt{true} \lor \mathrm{res} > 11 \doteq \mathtt{true}$ under which the type is violated.
\end{enumerate}
If we can find values for $x_i$ and $\mathrm{res}$ that satisfy the formula, we know that there are cases where this return type is violated.
Conversely, if this formula is unsatisfiable, we know that no such values can exist, and we can thus conclude that the return type is correct.
Our system is incomplete but sound: While we cannot always create formulas for which satisfiability implies an ill-typed expression, unsatisfiability \emph{does} imply well-typedness.
We omit the formal definition for type check formulas for brevity; more details can be found in the Bachelor's thesis~\cite{thesis} which this paper is based upon.

\section{Implementation}
Our modification~\cite{artifact} extends the framework's Constant Value Checker such that any type check that fails with the included Checker invokes our SMT type-checking instead to avoid false claims of ill-typedness.
Furthermore, we extend the framework's non-dependent value types with our dependent expression types.
\subsection{Overview}
The SMT value checking extension is integrated into the existing Value Checker, but disabled by default.
If the extension is enabled, type checking follows these steps:
\begin{enumerate}
	\item The Value Checker generates a control-flow graph for the given part of the code and runs its transfer function in an attempt to determine whether the code is well-typed using its standard syntactic rules.
	\item If any type checks fail with the syntactic rules, the Checker instead invokes the SMT value checking subsystem, which creates a type check formula and passes it to one or multiple SMT solvers.
	\item As soon as one of the invoked solvers returns a result, a successful or unsuccessful type-check result is returned.
    If the formula is satisfiable, the type check fails and the checker outputs further information about sample values of expressions in the code with which a type error occurs, which are gathered from the model that the SMT solver has found.
    If the formula is unsatisfiable, the type check is successful.
    If none of the SMT solvers terminate within the specified timeout limit, a timeout error is returned and the type check also fails.
\end{enumerate}
\subsection{Applying SMT value checking to Java}
The complexity of the Java language requires further considerations when creating type check formulas.
While mutability of local variables is dealt with using an SSA translation as mentioned before, each method call is assumed to produce an independent value due to potential side effects and is thus translated to a unique formula variable.
For arrays, we assume the value for an index to remain constant until a statement that could cause a side effect is evaluated (e.g. an assignment to a field or array or a method call).%
\footnote{This assumption is only valid in single-threaded applications.
The extension includes an option for enabling concurrent semantics, which disables this assumption.}
Furthermore, while syntactic sugar statements such as \mintinline{java}{switch} statements and ternary expressions are trivially translatable to regular \mintinline{java}{if} statements, \mintinline{java}{try}-\mintinline{java}{catch} blocks require special treatment.
Since there is generally little information as to when a specific exception could be thrown, we treat \mintinline{java}{catch} blocks as additional conditional branches with unknown conditions, mirroring the potential control flow in cases where an exception could be thrown.

Another intricacy comes from the wrapper types for primitives.
The extension, much like the base Constant Value Checker, treats primitives and their respective boxed types as equivalent and thus considers boxing and unboxing to be identity functions.
Although the boxed types can have \mintinline{java}{null} values, the Constant Value Checker assumes all values to be non-null for its analysis, since any operation on them will result in a \mintinline{java}{NullPointerException}.%
\footnote{Any boxed primitive whose value is known to be \mintinline{java}{null} (e.g. due to a conditional statement testing for nullness) is automatically assigned the bottom type \mintinline{java}{@BottomVal}, i.e. any such variable automatically serves as a valid instance of any value type.}
\subsection{Integration with SMT solvers}
\label{sec:implementation-smt-integration}
The SMT-enabled Value Checker uses JavaSMT~\cite{javasmt}, a library providing a common Java interface to create formulas for and call into various supported SMT solvers.
This library serves as a common interface to multiple SMT solvers and allows us to use all of its supported solvers without directly depending on any singular one of them.

The SMT value checking subsystem implementation currently only supports Java expressions of the integer types \texttt{byte}, \texttt{short}, \texttt{int}, \texttt{long} and \texttt{char} as well as the \texttt{boolean} type.
While boolean expressions directly map to formulas in the \emph{uninterpreted functions with equality theory},%
\footnote{In theory, one needs to distinguish between boolean \emph{terms} (which have values in the interpretation's domain and are not well-formed formulas themselves) and predicates and boolean operations on them (which constitute a valid \emph{formula}).
However, most SMT solvers consider expressions with boolean domain values to be equivalent to logical formulas, which allows us to use logical operands directly in terms.}
which is decidable~\cite{euf-decidability}, there are two common theories that could be used for integer expressions:
\begin{inparaenum}
	\item \emph{Integer theory}, which uses regular mathematical integers.
	      These do not directly match the semantics of the finite Java integers, as arithmetic operations on them do not overflow, and bitwise operations cannot be easily translated.
          Furthermore, for non-linear arithmetic, the integer theory is non-decidable~\cite{gödelincompleteness}.
	\item \emph{Bitvector theory}, which uses bitvectors of fixed sizes, which closely match Java integer semantics including overflows, and can also be used to model bitwise operations, but can lead to more timeouts than the integer theory.
\end{inparaenum}
To support all Java integer operations, we chose to use the bitvector theory for our extension.

\section{Evaluation}
We evaluate our implementation on the basis of a set of methods with Value Checker type specifications.
For this purpose, we use the \emph{Z3}~\cite{z3} solver.
Afterwards, we compare the runtime for default and SMT-enabled value checking and assess the overhead incurred by the SMT-solving layer.
\subsection{Decrease in false negative type-check results}
\label{sec:decrease-false-negatives}
\subsubsection{Test cases with non-dependent value types}
We first demonstrate the additional strength gained due to the SMT value checking subsystem with a set of test cases only making use of the existing non-dependent value types in the Value Checker, for which the regular implementation is unable to prove well-typedness.
\paragraph{Correlation tracking}
As stated in the introduction, the Value Checker's manual typing rules do not perform correlation tracking between expressions.
This often leads to larger losses of precision in the inferred types, such as in the statement \mintinline{java}{@IntVal(0) int res = x-x;}, for which the Value Checker is unable to prove correctness w.r.t the constant value type.
These problems are largely alleviated with the SMT type checking extension and its consideration of assignment contexts, with which it can prove tighter bounds for the results of calculations like the polynomial evaluation seen in Figure \ref{fig:value-checker-degree-7-polynomial}.
\begin{figure}
	\centering
	\begin{minipage}{0.95\textwidth}
		\begin{minted}[autogobble]{java}
        // The return type is the tightest correct range for this expression,
        // since it evaluates to 318 for x = -27 and to -77 for x = -16.
        // Without the SMT extension, the tightest bound known to the Checker
        // is @IntRange(from=-54449, to=77785).
        @IntRange(from=-77, to=318) long pol(@IntRange(from=-30, to=21) long x)
        { return ((x+30)*(x+20)*(x+10)*x*(x-10)*(x-20)*(x-30)) / 3000000; }
        \end{minted}
	\end{minipage}
	\caption{A 7th-degree polynomial for which the SMT extension is successfully able to prove the return value type.}
	\label{fig:value-checker-degree-7-polynomial}
\end{figure}
\paragraph{Precision losses due to limits of types}
In other cases, the Value Checker loses precision not due to loss of correlation, but due to the limited expressiveness of the value types.
Consider the test case in Figure \ref{fig:value-checker-range-precision-loss}.
\begin{figure}
	\centering
	\begin{minted}[autogobble]{java}
    @IntRange(from=5, to=10) int precLoss(@IntRange(from=-10, to=-5) int x,
            @IntRange(from=5, to=10) int y) {
        @IntRange(from=-10, to=10) int someValue = someCondition() ? x : y;
        return someValue > 0 ? someValue : -someValue; }
    \end{minted}
	\caption{A sample method in which the Value Checker does not persist enough information about a statement due to the limited expressiveness of the value type system.}
	\label{fig:value-checker-range-precision-loss}
\end{figure}
We know that the variable \texttt{someValue} holds a value $x \in [-10, -5] \cup [5, 10]$.
However, since we can only specify a single value type annotation, the most precise range we can specify is $[-10, 10]$,%
\footnote{We could \emph{specify} the type as \mintinline{java}{@IntVal({-10,-9,-8,-7,-6,-5,5,6,7,8,9,10})} instead.
However, the Value Checker replaces an \mintinline{java}{@IntVal} annotation with more than 10 values with an \mintinline{java}{@IntRange} annotation whose range encompasses all values of the array to prevent it from checking too many combinations of values when applying the transfer function~\cite{constvalchecker}.
Therefore, the Value Checker would end up using the same range as mentioned above.}
meaning we lose the fact that \texttt{someValue} never holds a value in the range $[-4, 4]$.
As the default Value Checker can only use type information for a variable, it therefore can no longer prove the fact that the returned value is within the range $[5, 10]$.
In contrast, the SMT extension can conclude that \texttt{someValue} is never assigned a value within $[-4, 4]$ using its assignment context.
\subsubsection{Dependent type evaluation}
In addition to extending the type checking process with SMT solving, we introduced new dependent value types into the value type system.
When specifying these types, SMT value checking \emph{must} be enabled in the Value Checker, as the syntactic type rules cannot handle these expression types.
Using the dependent value types, we can annotate various mathematical functions and prove properties about their return values.
Some samples that are provable with the SMT extension can be found in Figure \ref{fig:value-checker-dependent-type-samples}.
\begin{figure}
	\begin{minted}[autogobble]{java}
    @IntValExpr({"#1", "#2"}) int min(final int first, final int second) {
        return first >= second ? first : second; }

    @IntRangeExpr(from="#2", to="#3") int clamp(int input,
            // Only one of the following two annotations is necessary
            final @IntRangeExpr(to="#3") int min,
            final @IntRangeExpr(from="#2") int max) {
        if (input < min) { return min; }
        if (input > max) { return max; }
        return input; }
  \end{minted}
	\caption{Sample methods using dependent value types to specify additional properties about their return value.
		The well-typedness for these is provable with the SMT extension.}
	\label{fig:value-checker-dependent-type-samples}
\end{figure}

\subsection{Overhead evaluation}
\label{sec:overhead}
To gain an insight into how the SMT extension for the Value Checker affects the type-checking runtime, we first run the internal Value Checker test cases, which encompass 114 classes directly testing the Value Checker as well as 349 tests common to all Checkers testing the handling of various Java language features, on a system with an AMD Ryzen 5 5600X 6-core processor and 16 GB of memory using Z3 version 4.16.0 and the Temurin JDK version 25.0.3+9, running each configuration 5 times.~\cite{artifact}
Using the Value Checker without the SMT extension on the integrated tests, we get an average runtime of about 10 seconds and 844 ms.
When enabling SMT value checking, the overhead of SMT solving for failing type checks increases the average to about 12 s and 873 ms, amounting to an increase of about 18.7\%.
Additionally using the \texttt{-AsmtForceSmtChecking} option to force SMT value checking on every type check leads to a much larger increase to an average of 17 s and 731 ms, or 63.5\% above the non-SMT configuration, showing that the default inference rules remain relevant as a filter for more trivial cases in which invoking the SMT subsystem would lead to a significantly longer runtime.

For evaluating the runtime of dependent value type checks (which are not supported by the regular Value Checker), we use an additional set of test cases, which can be found in the \texttt{checker/tests/value-smt-evaluation} directory of the artifact~\cite{artifact}.
Overall, for type checks that successfully produce a result, the runtime remains under $2$ seconds.
However, whether a given type check ends up running into a timeout is not entirely obvious from the type-checked code.
Consider the method in Fig. \ref{fig:timeout-method}.%
\footnote{The syntax \texttt{\#2} refers to the 2nd parameter and is used in the Checker Framework to ensure that type information is preserved even in compiled code that does not contain variable names.}
\begin{figure}
  \centering
  \begin{minipage}{0.9\textwidth}
  \begin{minted}[autogobble]{java}
  @IntRangeExpr(from="0", to="#2 - 1")
  int mod(@IntRange(from=0) int val, final @IntRange(from=1) int x) {
    return val % x; }
  \end{minted}
  \end{minipage}
  \caption{A simple method which leads to a timeout when type checking.}
  \label{fig:timeout-method}
\end{figure}
Despite the rather simple nature of this statement, Z3 is unable to find a solution for the type check problem when using the bitvector theory for integers.
However, modifying the extension to make use of the Integer theory instead \emph{does} make the type check succeed rather quickly, showing that the strength of the extension could be potentially improved using the Integer theory.

\section{Conclusion}
In this paper, we have presented an extension to the Value Checker from the Java Checker Framework that strengthens type checking with an SMT solver integration.
We showed that by re-checking negative type check results with an SMT solver, we are able to reduce the number of false errors while preserving soundness at an average run-time increase of about $12.1\%$.
Furthermore, we introduced additional dependent value types to the Checker to extend the expressiveness of the type system.
\bibliographystyle{splncs04}
\bibliography{paper}
\begin{appendix}
    \newlist{funcoc}{enumerate}{1}
\setlist[funcoc]{font=\sffamily,label=\textbf{F\textsubscript{\arabic*}},ref=\textsf{F\textsubscript{\arabic*}},align=parleft}
\newlist{reuseoc}{enumerate}{1}
\setlist[reuseoc]{font=\sffamily,label=\textbf{R\textsubscript{\arabic*}},ref=\textsf{R\textsubscript{\arabic*}},align=parleft}
\section{Artifact}
The following functional outcomes are expected:
\begin{funcoc}
    \item The example code snippets in Section \ref{sec:decrease-false-negatives} show various scenarios in which SMT-enabled value checking is able to validate correct type annotations that the original Value Checker falsely deemed incorrect, alongside some samples with dependent type annotations that the original Value Checker did not support.
    \label{foc:decrease-false-negatives}
    \item Section \ref{sec:overhead} shows that the overhead for enabling SMT value checking with the Z3 solver on the existing Value Checker test cases in the framework leads to a runtime increase of about $18.7\%$, while forcing all value type checking results to be re-checked with the SMT solver increases the runtime over the regular (non-SMT-enabled) Checker by about $63.5\%$.
    \label{foc:integrated-tests-overhead}
    \item Section \ref{sec:overhead} also shows that the included dependent type code samples that do not time out have a type checking runtime of less than 2 seconds.
    \label{foc:dependent-tests-overhead}
\end{funcoc}
The following reusability claims are expected:
\begin{reuseoc}
    \item The artifact can be used to perform value type checking on other programs with value type annotations.
    \label{roc:type-checking}
\end{reuseoc}
\subsection{Quick start}
\begin{figure}
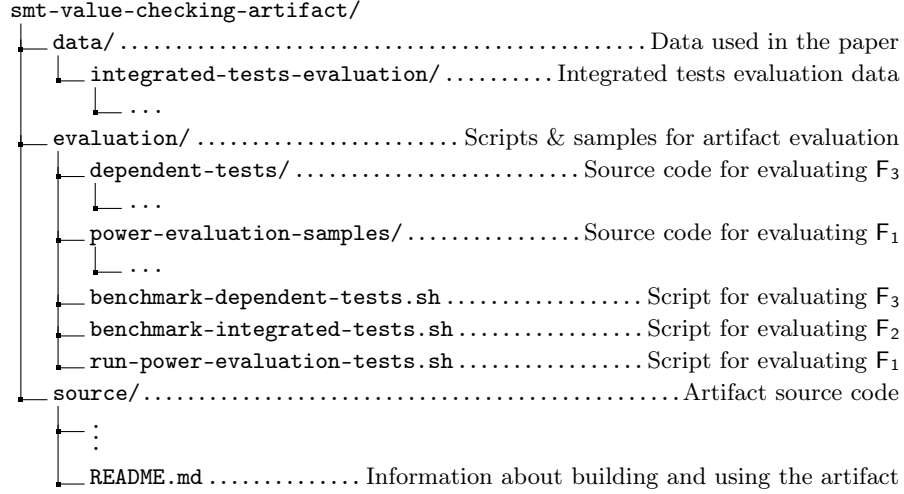

    \dirtree{%
      .1 smt-value-checking-artifact/.
      .2 data/\DTcomment{Data used in the paper}.
      .3 integrated-tests-evaluation/\DTcomment{Integrated tests evaluation data}.
      .4 \dots.
      .2 evaluation/\DTcomment{Scripts \& samples for artifact evaluation}.
      .3 dependent-tests/\DTcomment{Source code for evaluating \ref{foc:dependent-tests-overhead}}.
      .4 \dots.
      .3 power-evaluation-samples/\DTcomment{Source code for evaluating \ref{foc:decrease-false-negatives}}.
      .4 \dots.
      .3 benchmark-dependent-tests.sh\DTcomment{Script for evaluating \ref{foc:dependent-tests-overhead}}.
      .3 benchmark-integrated-tests.sh\DTcomment{Script for evaluating \ref{foc:integrated-tests-overhead}}.
      .3 run-power-evaluation-tests.sh\DTcomment{Script for evaluating \ref{foc:decrease-false-negatives}}.
      .2 source/\DTcomment{Artifact source code}.
      .3 \vdots.
      .3 README.md\DTcomment{Information about building and using the artifact}.
    }
    \caption{Overview of the artifact directory structure.
    The data referenced in the paper can be found in the \protect\path{data/} directory of the artifact.
    The \protect\path{evaluation/} directory contains data and scripts required for evaluating the functional outcomes of the paper.
    The \protect\path{source/} directory contains the source code of the Checker Framework fork with SMT value checking capabilities.
    }
    \label{fig:dirtree}
\end{figure}

The artifact repository contains an archive (\path{smt-value-checking-artifact.zip}) of the data and source code required for evaluating the artifact of the paper, alongside a Docker image (\path{docker-image.tar}) which includes the archive's files in the \path{/home/ubuntu/smt-value-checking/} directory (which is the working directory upon entering the container's shell) and contains the required dependencies for working with the artifact.
An overview of the artifact can be found in Figure \ref{fig:dirtree}.

Reviewers working with the artifact directly (without using the Docker image) need to have Git, Bash and the GNU coreutils installed alongside a Java Development Kit (JDK) of version 17 or 21.
Reviewers who wish to use the Docker image can import it using \texttt{docker image load -i docker-image.tar} and then create and run a container with the image using \texttt{docker run -it smt-value-checking}.

Before evaluating, reviewers should run \texttt{./gradlew assembleForJavac} in the \path{source/} directory of the archive files and ensure that the build completes successfully.
The build will produce the JAR files and native libraries required for running the forked Checker Framework in the \path{checker/dist/} directory of the source tree, which are used by the scripts in the \path{checker/bin/} directory.
The \path{checker/bin/javac} wrapper script can then be used in place of the regular Java compiler to compile files with value type annotations, performing type checking during the compilation process if the corresponding checker is enabled (using \texttt{-processor org.checkerframework.common.value.ValueChecker}).
The Docker image contains these artifacts in the \path{/opt/checkerframework/} directory, where the \texttt{javac} wrapper has been renamed to \texttt{cf-javac} to allow distinguishing between the JDK compiler and the wrapper script, as both are part of the \texttt{PATH}.
More information on the command-line flags related to SMT value checking can be found in the source tree's \path{README.md}.
\subsection{Functional evaluation}
\paragraph{Functional outcome \ref{foc:decrease-false-negatives}}
Running the \path{evaluation/run-power-evaluation-tests.sh} script will run the SMT Value Checker twice on the samples referenced in \ref{sec:decrease-false-negatives}, which can be found in the \path{evaluation/power-evaluation-samples/} directory: once with SMT value checking disabled, and once with SMT value checking enabled.
The first run will output various type checking errors referencing either supposedly incorrect types or the fact that dependent types are not supported without SMT value checking.
The second run will, in contrast, not output anything, indicating that all type checks succeeded.
\paragraph{Functional outcome \ref{foc:integrated-tests-overhead}}
Running the \path{evaluation/benchmark-integrated-tests.sh} script will run the integrated tests on the three different configurations five times as mentioned in Section \ref{sec:overhead}.
It will also write the runtimes of each run and each configuration into \path{evaluation/integrated-tests-results.csv} as a comma-separated values file.
Note that the test cases will always be run using the Gradle build tool in the source tree, since the test cases have more complex dependencies.
Runtimes are expected to differ from the results in the paper depending on the hardware used for evaluation, but the relative overhead of the two SMT configurations in relation to the non-SMT configuration should be similar.
\paragraph{Functional outcome \ref{foc:dependent-tests-overhead}}
Running the \path{evaluation/benchmark-dependent-tests.sh} script will compile the sample classes for testing dependent types referenced in Section \ref{sec:overhead} three times and log the time for each compilation in \path{evaluation/dependent-tests-results.csv} as a comma-separated values file.
Classes for which the compiler does not finish compilation within 1 minute are logged with a dash (-) instead.
As with \ref{foc:integrated-tests-overhead}, runtimes are expected to differ slightly depending on the hardware used for evaluation, but test classes without the \texttt{TimeoutCases} classes should compile within a few seconds, while test cases with the aforementioned suffix are expected to run into the timeout.
\subsection{Reusability evaluation}
This section demonstrates how to use the artifact of the paper for value type checking on custom Java classes with value type annotations (\ref{roc:type-checking}).
In the following, \texttt{<CF\_JAVAC>} will refer to the compiler wrapper script of the artifact.
Users of the Docker image can simply replace it with \texttt{cf-javac}, which is part of the image's PATH.
Users who built the artifact from source should replace it with the path to the \path{checker/bin/javac} file from the source tree.
Alternatively, the \path{checker/bin/} and \path{checker/dist} directories can also be copied together to a different directory, from which the \path{bin/javac} script can also be executed.

To perform value type checking on a custom Java file:
\begin{enumerate}
    \item Create a Java source file in any directory and declare a class inside of it.
    \item Import the \texttt{org.checkerframework.common.value.qual} package in the source file with \mintinline{java}{import org.checkerframework.common.value.qual.*;}
    \item Create methods that perform calculations on integers or booleans, and annotate potential parameters, fields and return types (if fitting) with value types using the corresponding annotation before the declared base type.
    Refer to Section \ref{sec:overview} for the available value types and their parameters and restrictions.
    \item Run the SMT Value Checker with the following command:
    \begin{center}
        \texttt{<CF\_JAVAC> -processor org.checkerframework.common.value.ValueChecker -AsmtValueChecking <source file>}
    \end{center}
    Replace \texttt{<CF\_JAVAC>} with the artifact's compiler wrapper and \texttt{<source file>} with the path to your source file.
    The artifact will run the Value Checker with SMT solving on the provided input, outputting any potential type errors, and will compile the class if no errors are found.
\end{enumerate}
Users may find it helpful to adapt or refer to one of the source files in the source tree's \path{checker/tests/value-smt-evaluation} directory as examples for programs with value type annotations.

\end{appendix}
\end{document}